\documentclass[12pt,preprint]{aastex}

\usepackage{natbib}
\usepackage{psfig}

\slugcomment{Submitted to ApJ, 29 October 2001; in original form 30 June 2001}
 
\shortauthors{McLaughlin et al.}
\shorttitle{Iron-group lines in GRBs}

\newcommand{\grb}{$\gamma$-ray burst}
\newcommand{\grbs}{$\gamma$-ray bursts}

\newcommand{\msun}{\mbox{$M_\odot$}}

\begin{document}


\title{Broad and shifted iron-group emission lines in \grbs\ \\
    as tests of the hypernova scenario}


\author{G. C. McLaughlin, R. A. M. J. Wijers, G. E. Brown}
\affil{Department of Physics and Astronomy, State University of New York,
    Stony Brook, NY 11794-3800}
\email{gail@tonic.physics.sunysb.edu, rwijers@mail.astro.sunysb.edu,
       epopenoe@nuclear.physics.sunysb.edu}

\and

\author{H. A. Bethe}
\affil{Floyd R. Newman Laboratory of Nuclear Studies, Cornell University,
    Ithaca, NY 14853, USA}

\begin{abstract}
In the hypernova/collapsar model of \grbs, it is natural that radiation
is emitted by the inner engine for some time after the burst. This has
been discussed as a possible source of the X-ray line emission observed
in some afterglows. We show here that the natural geometry of a hypernova
 --a source of radiation at the bottom of a deep funnel-- has
very significant consequences for the shape and central energy of the
observed emission lines: the lines acquire a very broad scattering wing
on the low-energy side and a characteristic second peak one Compton
wavelength away from the initial energy, due to the once- and twice-scattered
photons. We suggest that this explains the large width of the observed lines.
Furthermore, the downscattering lowers the central line energy (by up to
1\,keV in the source rest frame, before the lines become unrecognizable),
so that the observed line energies can
become consistent with originating from cobalt and nickel, as expected in
few-day-old supernova ejecta.
\end{abstract}


%%%%%%%%%%%%%%%%%%%%%%%%%%%%%%%%%%%%%%%%%%%%%%%%%%%%%%%%%%%%%%%%%%%%%%%%%%%%%%%%
\section{Introduction\label{intro}}

The discovery of \grb\ afterglows in X-ray \citep{Costa97B}, optical
\citep{Paradijs97}, and radio \citep{Frail97A} wavelengths greatly
increased the possibilities for studying the objects emitting this
radiation. 
This quickly led to 
the first measurement of a redshift \citep{Metzger97B} and the realization
that the relativistic fireball-blastwave model for their nature 
\citep{Rees92,Meszaros97} explained the afterglows very well
\citep{Wijers97B}. (For recent reviews, see Van Paradijs,
Kouveliotou, and Wijers \citeyear{Paradijs00}
and \citealt{Piran99}.) The phenomenal inferred 
emitted energy is at least that of a supernova; together with rapidly
variable $\gamma$-ray emission this
suggests a stellar-remnant origin.  Leading candidates
of this type are mergers of neutron stars and/or black holes 
\citep{Eichler89,Mochkovitch93, Janka99, Salmonson01}
or core collapses of massive stars \citep{Woosley93B,Paczynski98}. 
The latter are gaining favor
for long \grbs\ (duration over 2\,s \citep{Kouveliotou93};
these are the only ones for which afterglows have been found), for
a number of reasons: (i)
Some GRBs have shown evidence of association with a supernova, either
by a clear supernova found in their error box (GRB\,980425/SN\,1998bw;
\citealt{Galama98C}), or by late-time afterglow light curves showing a feature
consistent with a supernova. Clearest among the latter are GRB\,980326
\citep{Bloom99C} and GRB\,970228 \citep{Reichart99A,Galama00A}, though
other less clear cases have been claimed. Given the relative rarity of
these cases and the difficulty of detecting a supernova in the presence
of a bright afterglow, it is unclear whether this association holds 
more generally. (ii) The small spatial 
offsets between \grb\ counterparts and
their host galaxies (e.g., \citealt{Bulik99,Bloom99A,Bloom00B}) 
are hard to reconcile
with merging neutron stars, given their high expected space velocities and
long merger times, which should have them merge far away from where they
were born. (iii) The X-ray column density and extinction, as well as the
prevalence of strong MgI absorption lines in many optical
spectra of afterglows, indicate that \grbs\ often occur in dense regions
and behind much larger column densities than expected from random locations
in galaxies \citep{Galama01}. 
This suggests they occur amidst the large amounts of dust and
gas one expects to find in star forming regions. (iv) There is increasing
evidence that \grbs\ are associated with star formation on a more
global scale, both because the \grb\ rate as a function of redshift is
consistent with following the star formation rate 
\citep{Totani97,Wijers98,Krumholz98,Kommers00A,Schmidt01}
and because the average star
formation rate in \grb\ host galaxies is significantly above that of
samples of galaxies at the same redshift \citep{Fruchter01}.

An emission line has recently been found in some X-ray afterglows of \grbs\
\citep{Piro99A,Antonelli00,Piro00B}, whose 
energy is roughly consistent with Fe K$\alpha$ 
at the redshift of the host. This line has been adduced as evidence that 
the environment of the burst is heavily enriched in iron, and thus the result
of a recent supernova explosion. If this interpretation is correct, 
it would provide strong support for the connection between \grbs\ and 
stellar explosions. This makes the lines of considerable importance to
unveiling the nature of \grbs. In this paper, we investigate some problems
with the interpretation of the lines, and suggest a mechanism that may
help solve them. We begin by discussing models for the formation of iron
group lines in \grb\ afterglows and their difficulties in accounting for all
the observations (sect.~\ref{motiv}). Then we discuss our model for 
electron scattering of line photons in a hypernova funnel (sect.~\ref{model})
and the types of emission line profile that it predicts (sect.~\ref{resul}).
Finally, we discuss the relevance of our findings to current and future
observations and some possible complications
(sect.~\ref{discu}) and summarize our conclusions 
(sect.~\ref{concl}).

%%%%%%%%%%%%%%%%%%%%%%%%%%%%%%%%%%%%%%%%%%%%%%%%%%%%%%%%%%%%%%%%%%%%%%%%%%%%%%%%
\section{Motivation\label{motiv}}

The general thrust of models for the X-ray lines is that X rays from the
burst or afterglow irradiate cooler material, ionizing it almost completely.
Recombination of the material causes emission lines, and around 6--8\,keV
the iron group elements are the dominant contributors. The very large iron
line luminosity, about $10^{52}({\rm d}\Omega/4\pi)$ photons/s, 
sustained for a day or so, requires
$10^{57}({\rm d}\Omega/4\pi)$ emitted photons (${\rm d}\Omega$ is the
solid angle illuminated by the iron-line source).
If every atom emitted only once (e.g., if
the ionization happened in a flash, and every atom recombined only once, or if
the recombination time were of order the observed emission time), one
would need of order 50\,$({\rm d}\Omega/4\pi)$\msun\ 
of pure iron group material to take part. Therefore, one
needs a dense enough medium to have many recombinations per atom, but even then
detailed models often require the presence of large amounts of this material.
This gives rise to the claim that the irradiated medium is heavily enriched
in iron, and thus must have been enriched by a supernova quite recently.
The reported BeppoSAX WFC discovery of a transient absorption
feature in the prompt burst spectrum of GRB\,990705
\citep{Amati00} would, if typical of hypernovae, also set stringent
constraints on the amount of enriched ejecta and their distance from 
the \grb.

\begin{figure}
\psfig{file=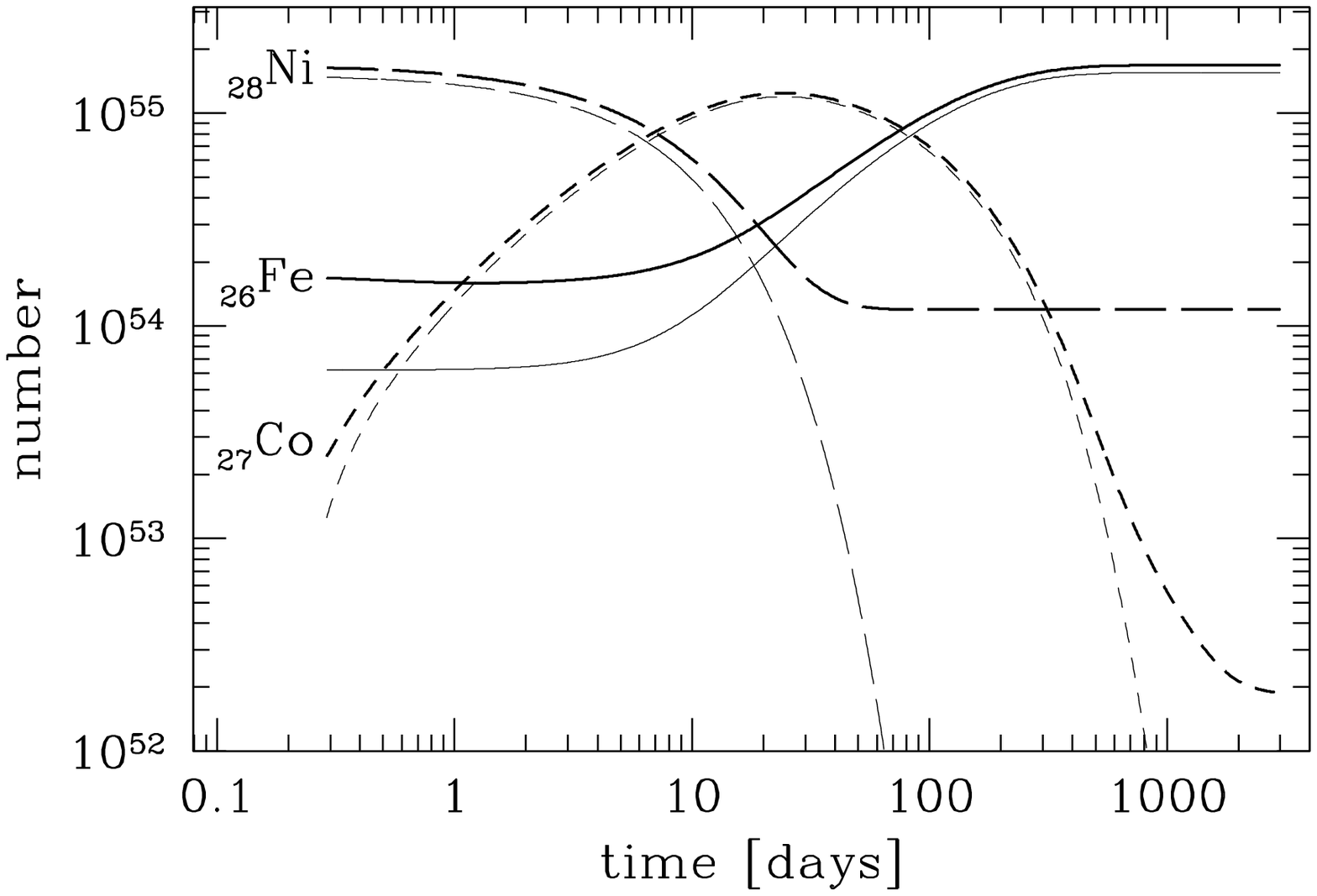,width=0.5\textwidth}
\caption{
Abundances of nickel, cobalt, and iron, as a function of time since the
supernova for a 40\msun\ star (model S40C, 16\msun\ He core) as modeled by
Woosley \& Weaver \citeyearpar{Woosley95B}. The thick lines give
the sum over all isotopes, the thin ones represent
only the $A=56$ isotope of each.
 \label{funfig1}}
\end{figure}

The models for the origin of the line come in two basic kinds. In one type
of model, the supranova, the medium is a shell ejected in an explosion
some months prior to the \grb\ \citep{Vietri01}. When the burst flux
hits the shell it emits its lines; as one needs many photons per atom,
the recombination time should be much less than the burst duration,
so the density needs to be high. However, the observed duration of the
line emission, about a day, is much longer. One can get this by
choosing the shell size to be about a light day, so that the duration of
the emission is set by the light crossing time of the shell. The emission
is naturally isotropic, so $10^{57}$ line photons must be produced.
Weak points
of this model are a few: first, the observed supernova `bumps' in \grb\
afterglow light curves
are consistent with the supernova going off at the same time as the
\grb, rather than the required few months before the burst.  Second,
the required time delay is created by having a supernova first make a
massive neutron star which is partly supported by rotation. When it slows
down, its centrifugal support gives way at some point and it collapses to
a black hole, giving a \grb. The problem with this may be that there
is little reason for the delay between supernova and \grb\ to be even
approximately constant, which in turn makes it hard to understand why
the iron shell should be about a light day across. Third, the lines have
thus far been identified as iron lines, and in cases with an independently
known redshift that does appear the best in terms of the observed line
energy of 7\,keV \citep{Piro99A,Piro00B}. 
However, a supernova produces primarily nickel, not
iron. In figure \ref{funfig1} we show the total abundance of iron,
cobalt, and nickel as a function of time since the supernova. Iron does
not become the most abundant element until three months after the supernova,
and it takes over half a year for it really to dominate.

The other type of model is different than that described above in
two respects: the irradiated
material is presumed to be closer to the source (e.g.,
\cite{Boettcher00}), most likely leftover
material from  the disrupted star. Also, the irradiation source is
long-lived. It could either be the afterglow itself or residual accretion
onto the compact remnant of the hypernova. The fact that the line flux
and afterglow flux evolve differently with time favors the latter
(e.g., in GRB\,000214; \citealt{Antonelli00}).  A slight variation on this
is the model by Rees and M\'esz\'aros \citeyearpar{Rees00}, who assume the
compact object emits a relativistic wind, which gives rise to heating and
radiation when it hits the disrupted star. This makes little difference,
since in both cases the net effect is that the surface of the ejecta
produces the line radiation due to bombardment with X rays. In this
model, the density of the line emitting material can be much higher,
and thus the number of recombination photons per atom is much larger.
Since also this model is naturally anisotropic and thus requires fewer
photons produced to begin with, 
much less iron-group material is needed, but enrichment by
supernova ejecta still seems required \citep{Boettcher00}. This model
does not require a delay between supernova and \grb, and thus is more
consistent with the observed supernova signatures in \grbs. However,
the lack of delay aggravates the line identification problem: in the
first few days the ejecta really are dominated by nickel, yet the line
energy as observed is usually more consistent with iron.

While both types of model thus have problems, we feel that the
disagreement in timing between the observed signs of supernovae in
a few \grbs\ and the predictions of the supranova model are
very hard to overcome.  In this paper, we investigate the possibility
of remedying the line energy problem in the context of hypernova models,
where we use the idea from Rees and M\'esz\'aros \citeyearpar{Rees00} that
we see a line emitted from the walls of a funnel in the progenitor star
that has just been created by the burst. The wall is energized by residual
emission of energy from the compact object formed during the \grb.

%%%%%%%%%%%%%%%%%%%%%%%%%%%%%%%%%%%%%%%%%%%%%%%%%%%%%%%%%%%%%%%%%%%%%%%%%%%%%%%%
\section{Model and calculations \label{model}}

We model the funnel and the scattering and emission within it as
follows: the funnel is taken to be a cone with some fixed opening angle,
$\theta_0$ (to be precise, $\theta_0$ is the angle between the cone axis
and the wall).  The wall is taken to have a high enough density that the mean
free path of photons within the wall material is negligible compared to
the cone size, so that all scattering within the wall takes place in a
skin layer of negligible depth. Absorption processes are parametrized
by assuming a fixed, energy-independent ratio for the scattering to
absorption mean free path. Photons are emitted isotropically from a ring
on the cone surface some fixed distance from the apex of the funnel. The
ratio, $h_{cone}$, of the total height of the cone to the emission height
is another parameter of the problem.

We use Monte Carlo calculations to compute the photon propagation and
energy change.
Since the scattering of the photons is by free electrons, the
angular information of scattering events must be retained in order to
calculate the direction of the outgoing photon and to determine the
photon energy loss.
The calculations are undertaken as follows.  A photon is emitted on the
surface of the cone, and its direction is chosen randomly in spherical
coordinates ($\phi,\cos \theta$). In this coordinate system, $\theta = 0$ 
points directly out of the cone and $\theta = \pi$ points directly down;
 $\phi= 0$ points to the axis of the cone.
Depending on the direction the photon may either escape entirely, cross the 
cone and enter the wall at a new location, or enter the wall at its 
current location.  One condition to describe the escape of the photon is
\begin{equation}
\label{eq:theta-escape}
\theta < \theta_0.
\end{equation}
The condition which describes whether the photon will cross the cone
is given by
\begin{equation}
\label{eq:crosscond}
\cos \phi > \tan \theta_0 / \tan \theta.
\end{equation}
If the
photon crosses the cone, the change in height of the photon is given by
\begin{equation}
\label{eq:deltaheight}
{\Delta h \over h} = 2 {(\cos \phi + \tan \theta_0 / \tan \theta) \over
\tan \theta / \tan \theta_0 - \tan \theta_0 / \tan \theta}.
\end{equation}
If the new height, $ h_{new} = h + \Delta h $ is greater than the 
total height of the cone, then the photon escapes.  
If $h_{new} < h_{cone}$ the photon will enter the cone wall at the
new point and we then need the direction of the photon at
the cone coordinates of the new point.  The angle $\theta$ is 
unchanged at this new position.  However, the 
coordinate system has rotated in $\phi$, since $\phi = 0$ always 
points back to the center of the cone.  The new $\phi$ is given by 
\begin{equation}
\phi_{new} = \pi - \arcsin \left[ {\sin \phi \over 1 + 
{\Delta h \over h}}  \right].
\end{equation} 

If the photon enters the wall, for bookkeeping purposes, rotation matrices
are used to describe its direction in coordinates where $\theta = 0$ 
parallel to the edge of the wall.  For convenience, we will refer to
 these as the wall coordinates.
  The wall coordinates are given in terms
 of the cone coordinates as
\begin{equation}
\cos \theta_{w} = -\cos \phi \sin \theta \sin \theta_0 +
\cos \theta \cos \theta_0;
\end{equation}
\begin{equation}
\sin \phi_{w} = \sin \phi \sin \theta / \sin \theta_{w}.
\end{equation}

Inside the wall, the photon scatters off an electron at a distance
randomly determined by the photon mean free path $\lambda_{scatt}$,
such that the probability of the photon traveling a distance $d$ before
scattering is $P_{scatt} \propto \exp(-d/\lambda_{scatt})$.  Given this 
distance $d$, the
photon has a probability of being absorbed of 
$P_{abs} = 1 - \exp(-d/\lambda_{abs})$.  In our calculations, we take 
$\lambda_{abs} = n \lambda_{scatt}$  and we examine several
different values of $n$, as discussed
in the next section.
Rather than throw away photons which fail the absorption test, 
we track a photon
intensity which begins as $I =  1$ and is multiplied by the absorption 
probability after each scattering.  Every photon which has not
escaped the cone by the time its intensity has diminished to less 
than a percent is discarded in our calculations. For 
$\lambda_{abs} \approx 100$, these photons have typically diminished in energy
so they are between 0.8--1.2\,keV.  If the photon is not absorbed,
we calculate the distance from the edge of the wall at which the
scattering occurs, using d and the
two angles in the wall coordinate system, 
$x = d \sin \theta_{w} \cos \phi_{w}$.

The scattered photon has a direction relative to the photon before scattering
which is chosen randomly according to the Thomson cross section 
$\sigma_{Th} \propto 1 + \cos^2 \theta_{out}$, where $\theta_{out}$ is 
the angle
of the photon coming out of the scattering event relative to the angle of
the photon going into the scattering event $\theta_{w}$.  The energy of the
photon is degraded according to this angle: $E + \delta E = E /[1 + (E / m_ec^2)
(1 - \cos \theta_{out})]$, where $m_e$ is the mass of the electron.  

In order to
determine whether the photon can escape the wall, $\theta_{out}$ and 
$\phi_{out}$ must be converted to the wall coordinate system:
\begin{equation}
\cos \theta_{wout} = - \sin \theta_{w} \sin \theta_{out} \cos \phi_{out}
+ \cos \theta_{w} \cos \theta_{out} 
\end{equation}
\begin{eqnarray}
\sin \phi_{wout} = && (\sin \theta_{w} \cos \phi_{out} \cos \theta_{w} 
\sin \theta_{out}  \cr
&& + \cos \phi_{w} \sin \phi_{out} \sin \theta_{out}   \cr
&& + \sin \phi_{w} \sin \theta_{w} \cos \theta_{out})/\sin \theta_{wout}  
\end{eqnarray}

Once the new photon direction is described in the wall coordinate system, 
we determine whether it escapes the wall. We again randomly choose a
distance traveled, $d$,  according to the mean free path.  We update the
distance from the edge of the wall to be $x + \Delta x$, where 
$\Delta x = d \sin \theta_{wout}  \cos \phi_{wout}$. We test to see if 
the photon's path has taken it outside the wall.  If not, we continue to
allow the photon to scatter in the wall until it leaves the wall
 or its intensity has decreased to less than one percent. Once the photon 
leaves the wall we use rotation matrices to describe it again in the
cone coordinate system.  It then either escapes the cone entirely 
or crosses to a new point on the cone and enters the wall at a new height,
calculated as starting with Eq. \ref{eq:deltaheight}, with the exception 
that the initial path length into the wall, $d$, is not chosen randomly.  It is
the path length from the photon's exit, diminished by the amount needed to
exit the wall.
  
In summary, the photon bounces around in the cone and in the walls of the
cone until it either escapes or its intensity becomes
negligible.  The intensities and energies of all escaping photons are 
recorded and used to generate the line profiles discussed in the next
section.

%%%%%%%%%%%%%%%%%%%%%%%%%%%%%%%%%%%%%%%%%%%%%%%%%%%%%%%%%%%%%%%%%%%%%%%%%%%%%%%%
\section{Results\label{resul}}

We present line
profiles calculated from various cone heights and opening angles and absorption
mean free paths.  We make contact with observation by convolving our profiles
with Gaussian functions and adding power law spectra.  We also examine the 
effect of the finite lifetime of various isotopes produced in a 
supernova-like
explosion. In particular we look at the time rate of change of nickel, cobalt 
and iron abundances and the effect on the line profiles.

Before presenting the results of the Monte Carlo simulations, 
it is worth pointing out that there is an analytic expression for the
average number of scatterings a photon will have on the way out of the cone, 
in a certain limiting case.  If the photon does not wander into the wall and
only stays on the surface, there is no absorption, and the funnel is
infinite, the average number of bounces that a photon will have on its way
out of the funnel is,
\begin{equation}
\bar{N} = (1  +  \cos \theta_0)/ (1 - \cos \theta_0),
\end{equation}
where $\theta_0$ is again the opening angle of the cone.  One
can see that for small cone opening angles the average number of 
scatterings is quite large.  For example for $\theta_0 = 10^\circ$, the
average number of bounces before escape is 130, approximately half
of which are bounces `backward', i.e., bounces which do not result in cone
crossings.  Reducing the cone height to $h_{cone} = 2$, but keeping the rest
of the simplifications, this number is reduced to 27.  
This is due to the 
long tail in the distribution of number of scatterings; e.g., in
the infinite cone limit with $\theta_0 = 10^\circ$,
the median number of scatterings is 89 (for a height of 2, the median is 16), 
whereas the mode is 0 for any opening angle.

\begin{figure}
\psfig{file=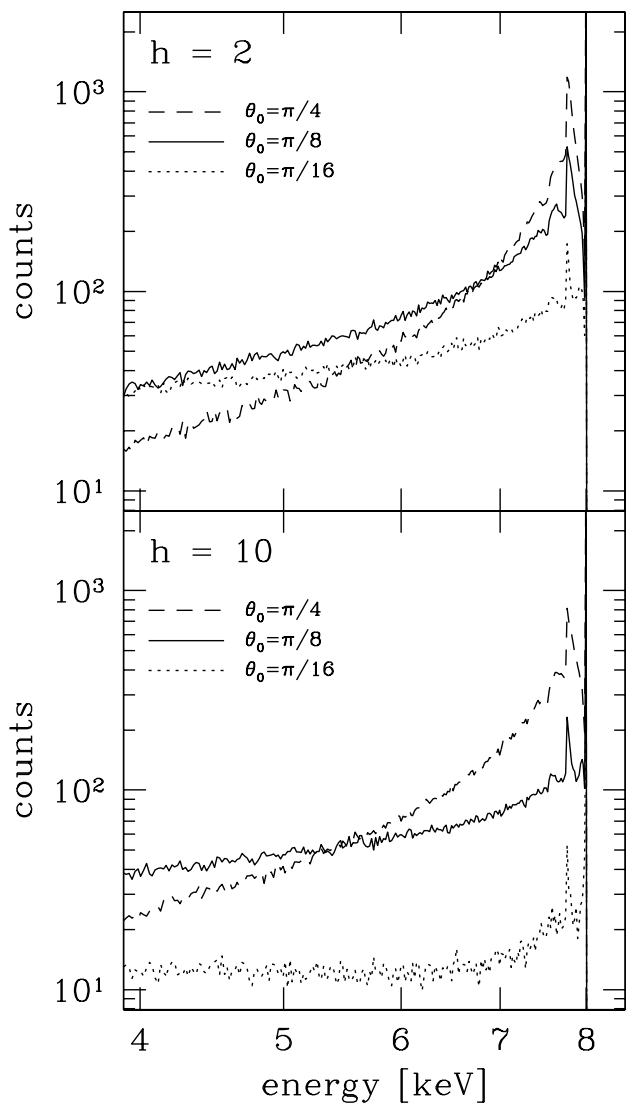,width=0.5\textwidth}
\caption{
Line profiles for cone heights of 2 (top) and 10 (bottom).  Photons of energy 
8\,keV are emitted at height 1.  In each panel we show the effect of 
decreasing opening angle: $\theta_0=\pi/4$ (dashed), $\theta_0=\pi/8$ (solid),
and $\theta_0=\pi/16$ (dotted). As the opening angle decreases, the peak
of few-times scattered photons near 8\,keV decreases, and the low-energy tail
becomes flatter. The depression is stronger for greater height at all opening
angles.
 \label{funfig2}}
\end{figure}

\begin{figure}
\psfig{file=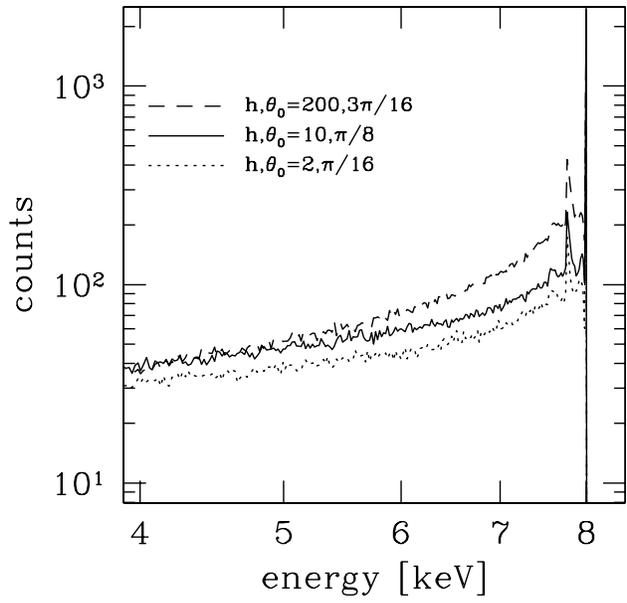,width=0.5\textwidth}
\caption{Increasing the opening angle of the cone has an effect which is 
very similar to decreasing the height, as illustrated by the similarity
of three cases: the top curve has $(h,\theta_0)=(200,3 \pi/ 16)$, the middle
one has $(10,\pi/8)$, and the bottom one $(2,\pi/16)$.
\label{funfig3}}
\end{figure}

\begin{figure}
\psfig{file=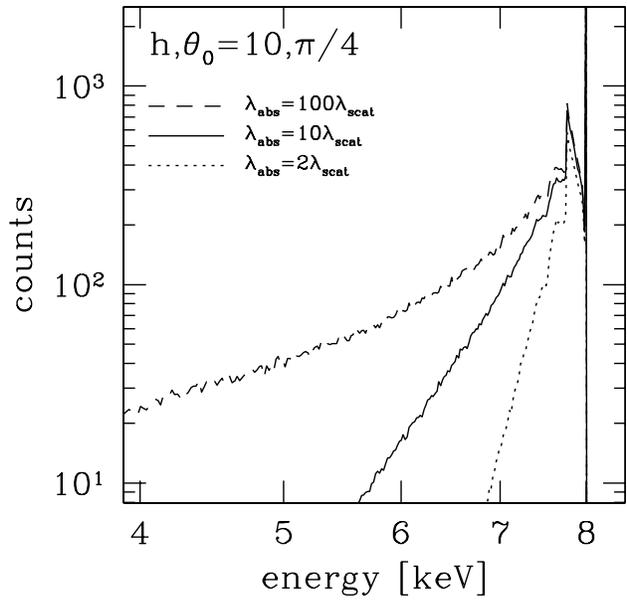,width=0.5\textwidth}
\caption{The effect of increasing the relative importance of
photon absorption.  The top curve is made with 
$\lambda_{abs} = 100 \lambda_{scatt}$, the middle one with $\lambda_{abs} = 
10 \lambda_{scatt}$, and the bottom one with 
$\lambda_{abs} = 2 \lambda_{scatt}$.  The opening angle of the cone is $\pi/4$ and the height is 10.
 \label{funfig4}}
\end{figure}

These long tails are also present in all of the line profiles presented here, 
which are calculated by the method described in the previous section.  
For example in Fig.~\ref{funfig2}
extended line profiles occur for
all the cone heights studied.  This figure shows results for
cone heights of $h_{cone} = 2$ and $h_{cone} = 10$.
The line photons are always emitted
from a height of 1, so the cone heights should be interpreted as ratios.
Both figures plot results for cone opening angles of $\theta_0 = \pi/16$, 
$\theta_0 = \pi/8$ and $\theta_0 = \pi/4$.  
One can see that increasing the opening angle has an effect which 
is similar to decreasing the height of the cone, and this is demonstrated 
in Figure \ref{funfig3}.  The tail length is strongly influenced  by the 
absorption mean free path, since the tail contains photons which have 
scattered many times.  Fig. \ref{funfig4} shows cases  
where the mean free path for absorption has been been reduced to
$\lambda_{abs} = 10 \lambda_{scatt}$, and 
$\lambda_{abs} = 2 \lambda_{scatt}$. Since the ratio is mostly a function
of temperature, with colder material being more absorbing, this 
illustrates the effect of changing the wall from hot to cold; for X-ray
photons, the wall retains considerable absorption opacity until the
temperature reaches the keV range. Mostly we
use a hot wall, with 
$\lambda_{abs} = 100 \lambda_{scatt}$, because a radiation bath that
suffices to ionize iron and nickel all the way to a hydrogen-like 
ion will also reduce the X-ray absorption opacity to very much below
that of a cold gas.

\begin{figure}
\psfig{file=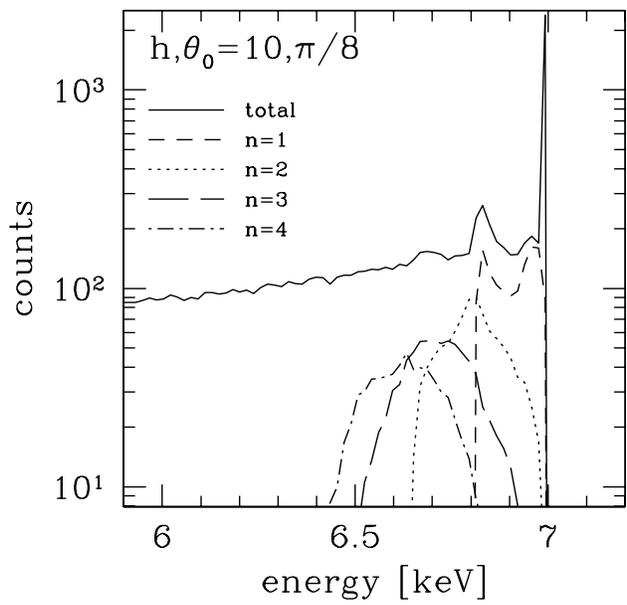,width=0.5\textwidth}
\caption{Decomposition of a line profile with a cone height of 10, and 
opening angle of $\pi/8$ into photons emitted with one, two, three and four 
bounces.  These are shown as the lower curves with number of bounces 
increasing from right to left. The overall line profile
is shown as the solid line.
\label{funfig5}}
\end{figure}

The curves which have the initial photons 
emitted from or close to the surface of
the wall have several features.  There is an initial spike at the energy
of the original line.  These are all of the photons which escape immediately 
without any scatterings.  There is also a pair of peaks
separated by about $\Delta E/E=2 E /m_ec^2$, with the higher-energy one
almost coincident with the escape spike.
The lower energy peak comes primarily from the photons 
which scatter once in the opposite direction of the opening angle and then 
scatter back, but also partially from twice scattered photons.  
Fig. \ref{funfig5} shows the decomposition of one of the line profiles into
one, two, three, and four times scattered photons.  The completely 
unscattered photons
are not plotted as a separate curve, since 
they simply comprise the initial spike.
The effect of the angular dependence of the Thomson cross section is
very clearly seen in the once scattered curve.  This is very similar to the
energies shown in \cite{Illarionov79} of photons scattered off free electrons,
but without any geometrical constraints such as in our study.

\begin{figure}
\psfig{file=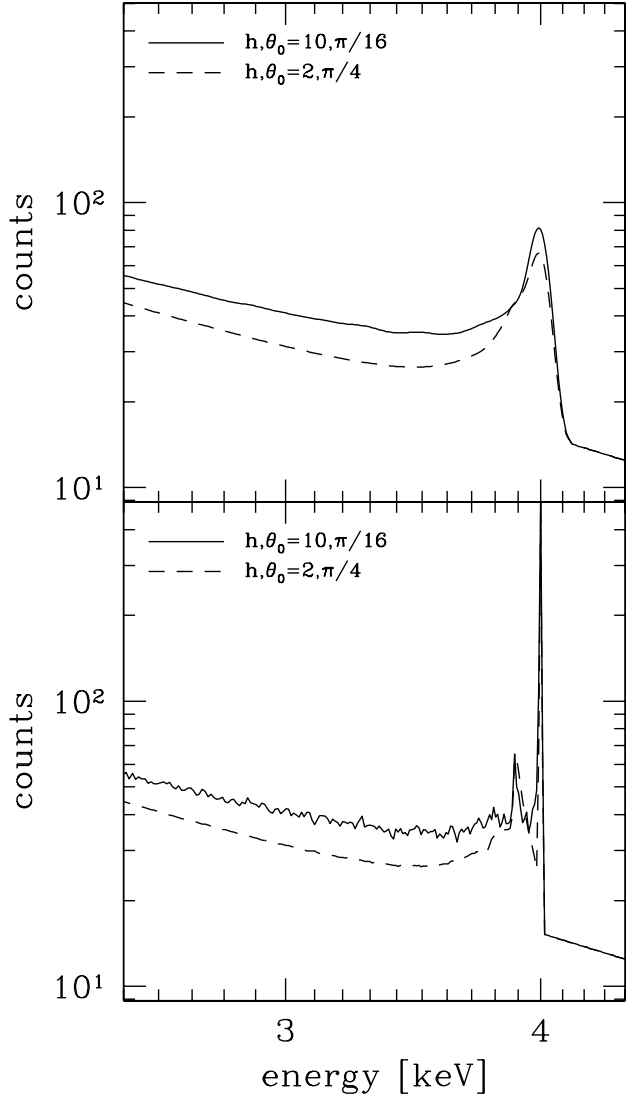,width=0.5\textwidth}
\caption{ 
Line profiles redshifted to $z=1$ and convolved with a 
Gaussian to simulate detector resolution.  A power-law spectrum has also
been added with ten times as many counts as in the narrow part of the
line, so the overall spectrum resembles that of the observed case of
GRB\,991216 \citep{Piro00B}.  The lower line shows the profile 
for a height of 2 and an angle of $\pi/4$. The upper line shows the profile 
for a height of 10 and an angle of $\pi/16$. The top panel is for a
resolution of 0.1\,keV around the line energy, mimicking the Chandra CCD,
and the bottom panel mimics the Chandra HETG with a resolution of 
0.003.
\label{funfig6}}
\end{figure}

We also show the lines convolved with Gaussians which mimic the spectral
resolutions of the CCD and the High Resolution Transmission Grating 
(HETG) of the  
Chandra X-ray observatory.  For
the former, we use a full width at half maximum of 0.1 keV and for the latter
a full width at half maximum of 0.0033\,keV; these values are appropriate
near $E=4$\,keV, where the lines are seen in our (observer) frame.
A power-law spectrum with
photon index $-2.2$ is added to
the profiles with ten times the number of counts (between 2 and 10\,keV) 
as in the peak of the line, to mimic the background against which the
line in the GRB\,991216 spectrum is seen.  We define the \lq peak\rq\ as 
the area within twice the full
width at half the initial peak height maximum, after convolution.  We have 
also given our line profiles a redshift of $z=1$, since recently detected
lines have occurred at roughly this redshift \citep{Piro00B}.  In figure 
\ref{funfig6} we show the effect of the CCD resolution and of the
high resolution grating.  The lower resolution smears out
most of the features of the line near the peak, but the large scattering
wing remains clearly visible.

\begin{figure}
\psfig{file=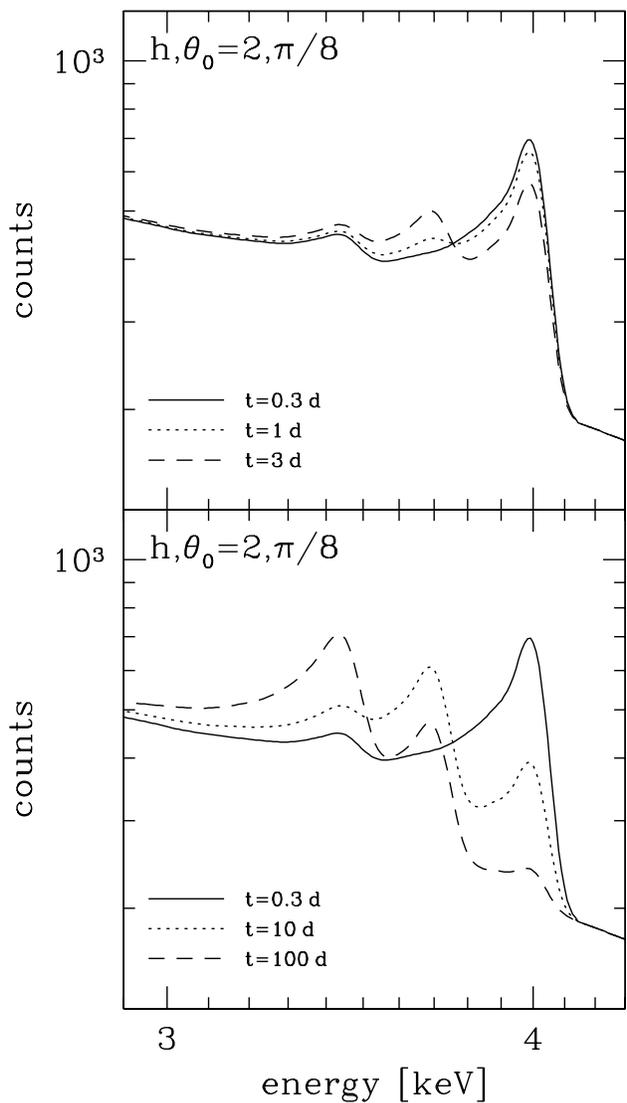,width=0.5\textwidth}
\caption{ 
The evolution of the redshifted line profile at CCD resolution
(0.1\,keV) as it decays 
from primarily nickel isotopes to iron, by way of cobalt.
In the top panel we show the line after 0.3 days (solid), 1 day (dotted),
and 3 days (dashed). The bottom panel is for delays of 0.3 (solid),
10 (dotted), and 100 (dashed) days.  In this figure the funnel 
height is 2 and the cone angle is $\pi/8$.
\label{funfig7}}
\end{figure}

The coarser CCD resolution however, is adequate for studying the time
variation of the signal due to decay of the nickel and cobalt 
isotopes.  In Figure \ref{funfig7} we show examples of 
these time profiles.  Here we have used the results of supernova nucleosynthesis
abundance calculations
of Woosley \& Weaver \citeyearpar{Woosley95B} as initial conditions
(Model S40C), and allowed the unstable isotopes to decay, keeping track
of all abundances.  Most of the change in the line profiles 
we present is due to the decay of the A~=~56 isotopes of nickel, cobalt and
iron.  We have used line energies of 8\,keV for nickel, 
7.4\,keV for cobalt, and 6.9\,keV for iron.  Nickel-56 has a relatively
short half life of 6.1 days as compared with Cobalt-56, which has a half 
life of 78.8  days. In figure \ref{funfig7} we show line profiles  
at 0.3, 1, 3, 10, and 100 days after the explosion.
Even shortly after the explosion, the line profile
clearly changes on a time scale of days as Nickel-56 decays to Cobalt-56.
At later times, one can clearly see the build-up of the 
iron abundance on time scales of months.    

One further curious point should be noted here: the decay of Nickel-56 is
almost exclusively by electron capture. This means that fully ionized Nickel-56
has a decay time many orders of magnitude longer than 
than neutral Nickel-56! The wall material we see is dominated by fully
ionized material, and therefore in principle Nickel could dominate the
observed emission even longer. Hydrogen-like Nickel-56 will decay only
half as fast as neutral Nickel-56, because only one of the capturable K
shell electrons is present. Cobalt-56 decays by both electron capture
and positron emission, so it will decay in fully ionized state, but
5 times more slowly than normal.  Of course, the ionized part of the wall is
very thin and thus contains almost no mass. Therefore, any amount of
mixing in the material that makes up the funnel will cause a given atom
to spend little time in the fully ionized state, and thus the influence
of ionization on the decay rate may be small in practice.

%-------------------------------------------------------------------------------
\subsection{Connection with real sources\label{resul.real}}

Thus far, we have phrased our results in terms of the minimal number of
(mostly dimensionless) parameters and thus kept them general. However,
we do primarily have the physical setting of a hypernova-GRB in mind,
so we briefly sketch a set of numbers to which our model applies, that
represent the likely situation in the aftermath of a GRB powered by
a collapsing massive star. It is very like the situation sketched by
\citealt{Rees00}. In order to have the required 10$^{52}$ photons/s in 
the observed line, we require perhaps 10$^{53}$ ph/s emitted above the
ionization threshold. This in turn implies an X-ray luminosity of
$L_{\rm X}\sim10^{45}\beta$ erg/s, where $\beta>1$ converts the ionizing 
luminosity into the total source luminosity.  
This is well in excess of the
Eddington luminosity of a stellar-mass black hole, and thus cannot be
emitted by the vicinity of the black hole in photons. This poses
no problems to the model because, as suggested by \citealt{Rees00},
it can be emitted as a particle wind and converted to X rays in shocks
near the funnel wall.  

If we take $h\sim2$, and
take the total funnel size to be of order the initial size of the star
($2\times10^{11}$\,cm), then the line photons will be primarily generated
at $r\sim10^{11}$\,cm. This implies a very high local X-ray flux at
the reprocessing site: $F_{\rm X}=10^{22}\beta$\,erg\,cm$^{-2}$s$^{-1}$.
Under such circumstances, the temperature in the funnel wall will be set
by ionization/recombination physics to a value a few times less than
the ionization threshold energy, which in this case is about $10^8$\,K.
The density in the funnel wall will be set by the fact that the wall
gas pressure must balance the impinging radiation pressure, which
implies $n_{\rm e}=10^{19}\beta$\,cm$^{-3}$. With this, we can assess the 
great importance of radiative ionization in the wall: the ionization
parameter, $\xi=L/nr^2=10^4$, indicating even iron-group elements are
largely ionized. It could even be so high that Thompson scattering
becomes important: the scattering mean free path is only 10$^5$\,cm,
illustrating that the wall is indeed very thin. Within this layer,
the electron temperature could become dominated by scattering, and set
to the Compton temperature, $kT_C\sim E_\gamma/4$, which is likely
of order 1\,keV for a typical X-ray spectrum. (Note that whether it is set
by Compton scattering or ionization balance, the temperature is well below
the line energy, justifying our approximation of a cold wall, in which
only Compton downscattering of the photons is important. Also, the
thermal width of any iron-group lines will be of order 1\,eV, negligible
relative to the scattering width.) 

\begin{figure}
\psfig{file=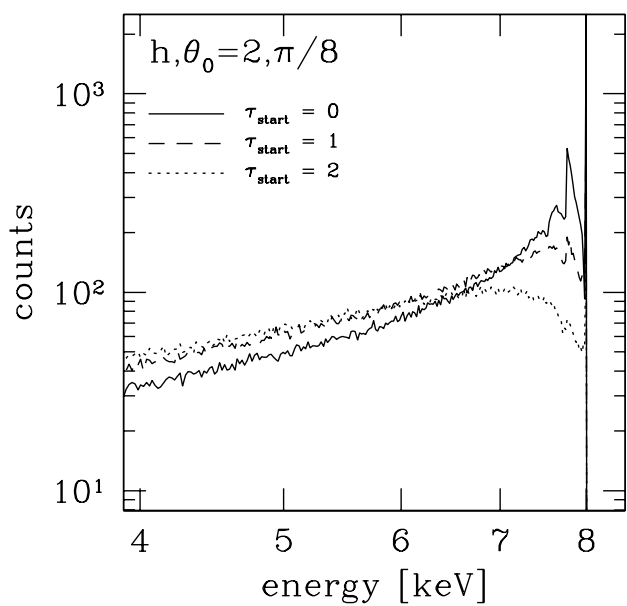,width=0.5\textwidth}
\caption{
The effect of  emission of the line photons at finite
scattering optical depth.
The solid curve is for photons initially emitted at the surface of the
wall.  The dashed line is for photons emitted at an optical depth of 1,
and the dotted line is for those emitted at an optical depth of 2. The
initial spike heights for the three curves are in a ratio of 22:4:1.
The plot was made for an opening angle of $\pi/8$ and a cone 
height of 2.
\label{funfig9}}
\end{figure}

The high degree of ionization of nickel means that the mean free path due
to the tiny fraction of hydrogenlike Ni atoms can be orders of magnitude
smaller than the scattering mean free path. This implies (Fig.~\ref{funfig9})
that the line may easily be scattered to a width that effectively makes
it invisible. Thus, the transient nature of the observed lines, as well
as their total absence in some GRBs, might well be due to too high
an ionizing flux instead of a too low one.

%%%%%%%%%%%%%%%%%%%%%%%%%%%%%%%%%%%%%%%%%%%%%%%%%%%%%%%%%%%%%%%%%%%%%%%%%%%%%%%%
\section{Discussion\label{discu}}

Our study of the influence of the funnel in hypernovae on emerging X-ray
lines shows that the shape of the lines is very strongly affected by
scattering off the funnel walls. This is true even if the funnel has 
an opening angle as large as 45 degrees and the line photons are produced
halfway between the bottom and top of the funnel. Quite generically, 
the lines can be broadened to 0.5-1\,keV FWHM, and their centers shifted
down in energy by up to 1\,keV before they become too wide to recognize
as lines. An important characteristic is that the
line remains fairly sharp on the blue side, with a broad red wing being
the main reason for the increased width and line-center shift.
If the lines are shifted down even
further, then they tend to become so smeared out that they may escape
detection against a background power-law spectrum from the source.

\begin{figure}
\psfig{file=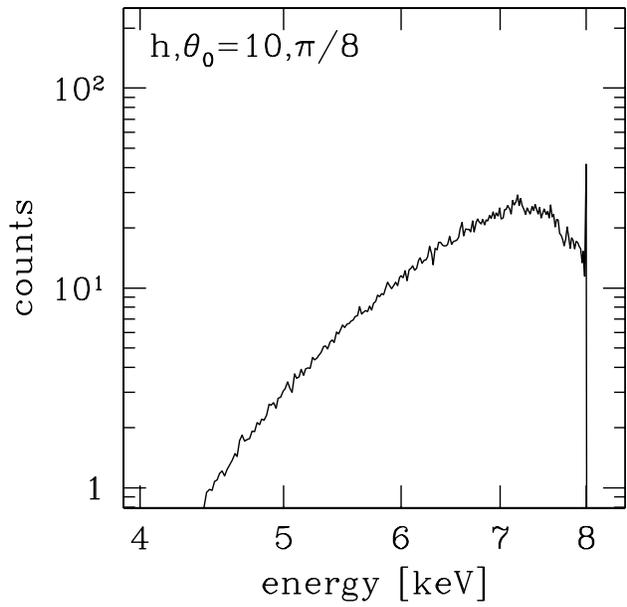,width=0.5\textwidth}
\caption{
The shift of an 8\,keV line to 7\,keV in a cone with $h=10$ and 
$\theta_0=\pi/8$.  The 
photons were emitted at an optical depth of 2 and the ratio of scattering
to absorption is 10. The shift of the line peak is due to the combined effect
of starting at finite optical depth, which depresses the blue side, and
absorption of the tail, which depresses the red side.
\label{funfig8}}
\end{figure}

In Figure \ref{funfig8} we show an example of how a line may appear to
become redshifted due to Compton scattering.  In this figure, 8 keV nickel
lines were emitted at an optical depth of 2 and 
$\lambda_{scatt} = 10 \lambda_{abs}$, from a cone with
$h=10$ and $\theta_0=\pi/8$.  The peak of this line occurs at 
around 7 keV, which happens to be the same energy as iron K$\alpha$ lines.
We see that absorption effects,
both as the photons make their way out of the wall and also on the tail
of the distribution conspire to create a peak about 1 keV below the 
original line energy.

The shape of the lines in the case of perfect scattering (i.e., no photons
are lost by absorption) is mostly set by the direct-escape probability
from the emission point. Since this probability reflects a combination
of the funnel opening angle and the funnel height relative to the emission
point, it follows that those two parameters are very difficult to 
disentangle from realistic data. The observed line wing is quite sensitive
to absorption, though: as the importance of absorption is increased, most
photons never escape the funnel. Since the escaping ones are always those
which have scattered few times, their energies have not changed very much.
As is illustrated by figures \ref{funfig2}--\ref{funfig5}, the lines always
have very extended wings in the perfect-scattering case, even if the
cone is wide and the escape probability high: there is always a significant
number of photons that scatter many times. The only exception to this is
absorption, which ensures that no photons scatter many times. Therefore,
narrowness of the line core due to wide funnels can be distinguished from
narrowness due to absorption by inspecting the red wing of the line.

A most interesting feature of the profiles is the structure near the 
original line energy: the escape peak, a sign of unscattered flux, and
two or three bumps near it that represent the 1--3 times scattered photons.
The bumps are very clear signs of a Compton scattering origin of the line,
and can only be seen with grating resolution. Unfortunately, their presence
is not guaranteed, as can be seen in figure \ref{funfig9}.
Rees and M\'esz\'aros \citeyearpar{Rees00} showed that close to the 
compact object, the required X-ray flux is so large that the ionized 
surface layer of the funnel becomes optically thick to Thomson scattering
(since the atomic edge cross section for nickel or 
iron is more than 100 times the
Thomson cross section, this requires extreme ionizing fluxes). This means
that a typical line photon is not created at the wall surface, but at a
few scattering optical depths into the wall. As a result, the number of
photons that escape with few scatterings becomes minimal very quickly,
removing the typical signatures of Compton scattering. (Note that 
especially for narrow funnels, the escape direction from the cone
is nearly perpendicular to the wall normal, so even a small optical depth
into the wall reduces the direct-escape probability to nearly zero.)

\begin{figure}
\psfig{file=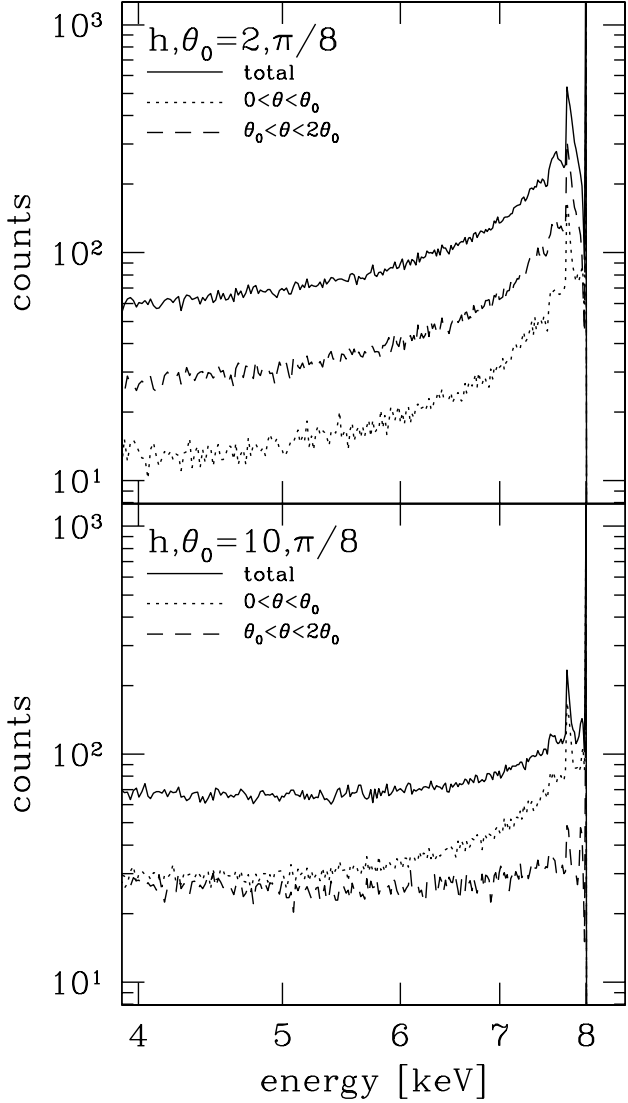,width=0.5\textwidth}
\caption{
The variation in intensity and shape of the line profile with
emission angle (or the observer viewing angle relative to the cone
axis), for an opening angle of $\pi/8$ and cone 
heights of 2 (top) and 10 (bottom). The top curve (solid) in both panels is
the sum over all emission angles. The dashed curve
represents the photons emitted with  angles between $\theta_0$ and 2$\theta_0$,
and the dotted one those emitted between between $0$ and $\theta_0$.
Note how the intensities of the on-axis and off-axis components are reversed 
between the shallow cone (top) and the deep cone (bottom).
\label{funfig10}}
\end{figure}

Another way in which the line shape can be affected is by the angle of
exit. Especially the flux that escapes the funnel from near the top,
at an angle greater than the cone opening angle, $\theta_0$, has a markedly
different spectrum than the total flux: the direct escape spike
and the back-scatter spike (i.e., the red spike of once-scattered photons
plus the peak of the twice-scattered ones) are markedly smaller
(figure \ref{funfig10}).

%%%%%%%%%%%%%%%%%%%%%%%%%%%%%%%%%%%%%%%%%%%%%%%%%%%%%%%%%%%%%%%%%%%%%%%%%%%%%%%%
\section{Conclusion\label{concl}}

We have shown that the funnel geometry associated with a hypernova
naturally produces some features of the emission lines observed in
afterglows. Especially the large width of the line seen in GRB\,991216
is a natural result of the large scattering wings induced by repeated
Compton recoils as a photon bounces around many times in the funnel
prior to escape. Therefore, it is not necessary to invoke very
high velocities of dense gas clumps or shells to explain the wide lines.
At the same time, this causes the center energy of the line to be lowered
by large amounts, depending on the funnel depth and opening
angle. Widths of 0.5--1\,keV, as observed, can be obtained reasonably well.
For significantly greater widths the line becomes unrecognizable, 
however.

The far red wing of the line provides a diagnostic of the relative
importance of scattering and absorption opacities in the funnel wall
and thus, of the temperature of the wall upon which this ratio
chiefly depends. At grating resolutions, one can furthermore recognize
sharp spikes in the line due to the photons that have scattered 0--2
times prior to escape. The magnitude of these lines is a measure of the 
escape probability from the bottom of the funnel (a combination of the
opening angle and the depth).

The shape of the lines is further affected by the composition of the
ejecta. Contrary to most earlier papers on the subject, 
the ejecta cannot predominantly emit iron lines unless they are older
than half a year. Any younger ejecta are dominated by nickel and/or 
cobalt. The only other time at which the ejecta are strongly dominated
by only one iron-group element is the first 1--2 days, but then the
dominant element is nickel, of which the K$\alpha$ line is at 8\,keV.
The discrepancy between this and the observed line energy of 7\,keV
can be explained by Compton downscattering.

In short, we feel that there is evidence in the lines of scattering origin
in a hypernova funnel. Future high-resolution spectra (with long exposure
times) will easily confirm or deny this proposition. If confirmed, the shape
of the lines provides direct diagnostics of the properties of the
hypernova remnants in \grbs.

\acknowledgments

GCM and GEB are supported by the U.S. Department of Energy under grant
DE-FG02-88ER40388. RAMJW is supported in part by NASA (award no.\ 21098).

%%%%%%%%%%%%%%%%%%%%%%%%%%%%%%%%%%%%%%%%%%%%%%%%%%%%%%%%%%%%%%%%%%%%%%%%%%%%%%%%
%   references
%%%%%%%%%%%%%%%%%%%%%%%%%%%%%%%%%%%%%%%%%%%%%%%%%%%%%%%%%%%%%%%%%%%%%%%%%%%%%%%%
%\bibliographystyle{astroshort}
%\bibliography{xmymoshort,x65,x70,x75,x80,x85,x90,x95,x00,x05}

\clearpage

%%%%%%%%%%%%%%%%%%%%%%%%%%%%%%%%%%%%%%%%%%%%%%%%%%%%%%%%%%%%%%%%%%%%%%%%%%%%%%%%
%   figure captions and figures
%%%%%%%%%%%%%%%%%%%%%%%%%%%%%%%%%%%%%%%%%%%%%%%%%%%%%%%%%%%%%%%%%%%%%%%%%%%%%%%%

\clearpage

\clearpage

\clearpage

\clearpage

\clearpage

\clearpage

\clearpage

\clearpage

\end{document}